\documentclass[aps,prb,reprint,groupedaddress]{revtex4-1}

\usepackage{graphicx} 
\usepackage{textcomp} 
\usepackage{hyperref} 
\usepackage{ifthen} 
\usepackage{xifthen} 
\usepackage{color} 
\usepackage{soul} 
\usepackage{amsmath} 

\newcommand{\fig}[2][]{%
\ifthenelse{\isempty{#1}}
{Fig.~\ref{#2}}
{Fig.~\ref{#2}#1}
}

\begin{document}

\title{Coherent control of a nanomechanical two-level system}

\author{Thomas Faust}
\author{Johannes Rieger}
\author{Maximilian J. Seitner}
\author{J\"org P. Kotthaus}
\author{Eva M. Weig}

\affiliation{Center for NanoScience (CeNS) and Fakult\"at f\"ur Physik, Ludwig-Maximilians-Universit\"at, Geschwister-Scholl-Platz 1, M\"unchen 80539, Germany}

\maketitle

{\bf The Bloch sphere is a generic picture describing a coupled two-level system and the coherent dynamics of its superposition states under control of electromagnetic fields\,\cite{Nielsen2000}.
It is commonly employed to visualise a broad variety of phenomena ranging from spin ensembles\,\cite{bib:Vandersypen2005} and atoms\,\cite{bib:Haroche2006} to quantum dots\,\cite{bib:Hanson2008} and superconducting circuits\,\cite{You2011}.
The underlying  Bloch equations\,\cite{bib:Bloch1946a} describe the state evolution  of the two-level system and allow characterising both energy and phase relaxation processes in a simple yet powerful manner\,\cite{bib:Vandersypen2005,bib:Yafet1963,bib:Burkard2004}.

Here we demonstrate the realisation of a nanomechanical two-level system which is driven by radio frequency signals.
It allows to extend the above Bloch sphere formalism to nanoelectromechanical systems.
Our realisation is based on the two orthogonal fundamental flexural modes of a high quality factor nanostring resonator which are strongly coupled by a dielectric gradient field\,\cite{Faust2012a}.
Full Bloch sphere control is demonstrated via Rabi\,\cite{bib:Rabi1937}, Ramsey\,\cite{bib:Ramsey1950} and Hahn echo\,\cite{bib:Hahn1950} experiments.
This allows manipulating the classical superposition state of the coupled modes in amplitude and phase and enables deep insight into the decoherence mechanisms of nanomechanical systems.
We have determined the energy relaxation time $T_1$ and phase relaxation times $T_2$ and $T_2^\ast$, and find them all to be equal.
This not only indicates that energy relaxation is the dominating source of decoherence, but also demonstrates that reversible dephasing processes are negligible in such collective mechanical modes.
We thus conclude that not only $T_1$ but also $T_2$ can be increased by engineering larger mechanical quality factors.
After a series of ground-breaking experiments on ground state cooling and non-classical signatures of nanomechanical resonators in recent years\,\cite{OConnell2010,Teufel2011,Chan2011,PhysRevLett.108.033602,2012arXiv1206.5562P}, this is of particular interest in the context of quantum information processing\,\cite{Nielsen2000,bib:Ladd2012} employing nanomechanical resonators\,\cite{bib:Stannigel2010,2012arXiv1211.4456R}. 
}


While the dynamics of a two-level system under the influence of a pulsed external electromagnetic field was observed in atomic and nuclear spin physics decades ago, a mechanical analogon to such a system remained elusive for a long time.
Only recently, coherent exchange of energy quanta between a mechanical and an electrical mode was achieved: In 2010, O'Connell et al.\,\cite{OConnell2010} managed to control the swapping of a single quantum of energy between a qubit and a mechanical resonator, while Palomaki et al.\,\cite{2012arXiv1206.5562P} demonstrated the temporary storage of itinerant microwave photons in a mechanical resonator in 2012.
At the same time, several approaches were employed to achieve purely mechanical resonant coupling either between separate resonators\,\cite{perisanu:063110,APEX.2.062202,PhysRevB.79.165309} or different modes of the same resonator\,\cite{kozinsky:253101,Faust2012a} in the classical regime.
So far, the pulsed coherent control of the system was prevented by weak coupling, low quality factors or the lack of a sufficiently strong and fast tuning mechanism.

\begin{figure}
\includegraphics{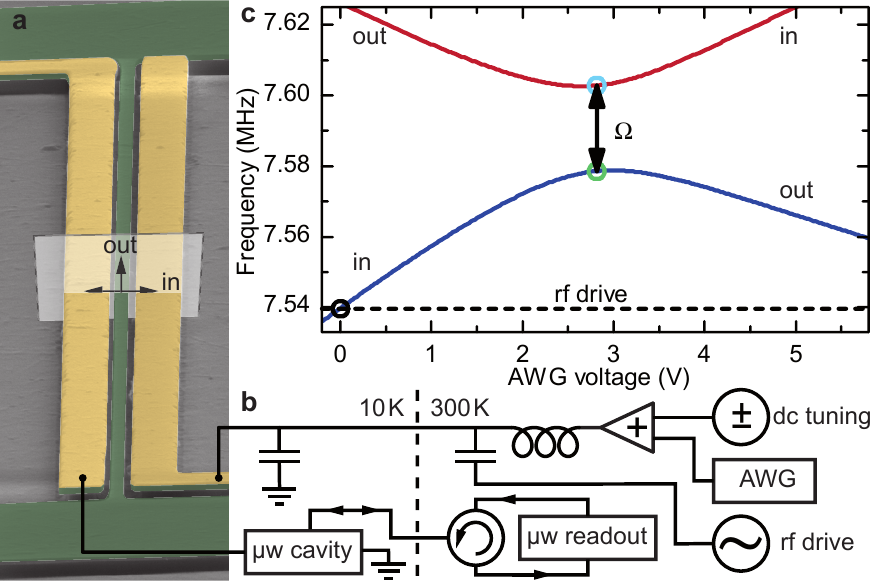}
\caption{\label{system}\bf Nanoelectromechanical system. a\sf , SEM micrograph showing oblique view of the 50\,\textmu m long silicon nitride beam (green) and the adjacent, 1\,\textmu m wide gold electrodes (yellow), processed on top of the SiN. \bf b\sf , Electrical setup: the output of the arbitray waveform generator (AWG) and a dc tuning voltage are added and combined with the rf drive via a bias-tee. The second capacitor acts as a bypass providing a \textmu w ground path for the microwave detection\,\cite{Faust2012}, which is connected to the other electrode. \bf c\sf , Resonance frequencies of the out-of-plane (out) and in-plane (in) mode of the resonator controlled by the AWG voltage at a constant dc tuning voltage of -15\,V. The black circle marks the initialisation state at 0V and the frequency of the rf drive, while the green and blue circles correspond to the lower and upper state of the two-level system, respectively, separated by the frequency splitting $\Omega$.}
\end{figure}

We present the successful implementation of a purely mechanical two-level system with coherent time-domain control (see also the experiments independently performed at NTT using parametric coupling\,\cite{Okamoto2012}).
To this end, we use a 250\,nm wide and 100\,nm thick, strongly stressed\,\cite{PhysRevLett.105.027205} silicon nitride beam resonator with a length of 50\,\textmu m dielectrically coupled to a pair of electrodes used for detection\,\cite{Faust2012} as well as actuation and tuning\,\cite{rieger:103110}.
The two fundamental flexural modes of the mechanical resonator oscillating in the out-of-plane and in-plane direction (see \fig{system}) are coupled by cross-derivatives of the strong inhomogeneous electric field generated between the electrodes\,\cite{Faust2012a}.
A constant dc voltage of -15\,V is used to dielectrically tune the system close to the resulting avoided crossing, while the signals generated by an arbitrary waveform generator (AWG) enable time-resolved control vicinal to the anticrossing (see \fig[b,c]{system}).
Both voltages are added and combined with the rf actuation of the beam via a bias-tee and applied to one electrode.
The other electrode is connected to a 3.6\,GHz microstrip cavity, enabling heterodyne detection of the beam deflection\,\cite{Faust2012} after addition of a microwave bypass capacitor at the first electrode\,\cite{rieger:103110}.
These components as well as the mechanical resonator are placed in a vacuum of $\leq10^{-4}$\,mbar and cooled to $10.00\pm0.02\,$K to improve the temperature stability as well as cavity quality factor. The microwave cavity is interfaced to the readout with a single coaxial cable and a circulator.

When the system is driven by an external white noise source and the AWG output voltage is swept, the avoided crossing of the two modes shown in \fig[c]{system} can be mapped out, exhibiting a frequency splitting $\Omega=24,249\pm 4\,$Hz.
With a quality factor $Q=\frac{f}{\Delta f}\approx 2\cdot10^{5}$ and a linewidth of $\Delta f\approx 40\,$Hz at the resonance frequency $f$, the system is clearly in the strong coupling regime of $\Delta f \ll \Omega$.
For all measurements discussed in the following, an rf drive of -59\,dBm at 7.539\,MHz, resonantly actuating the beam at an AWG voltage of 0\,V, is applied, which initialises the system in its in-plane mode (see black circle in \fig[c]{system}).
A 1\,ms long, adiabatic voltage ramp up to 2.82\,V brings the state to the point of minimal frequency splitting $\Omega$ between the coupled modes.
Here, the system dynamics is described by two hybrid modes formed by the in-phase and out-of-phase combinations of the fundamental flexural modes.
The adiabatic ramp thus transforms all the energy of the in-plane mode into the lower hybrid state, such that the two-level system, consisting of the two hybrid modes, is prepared in its lower state.
As the drive frequency remains constant (dashed line in \fig[c]{system}), the beam is no longer actuated and its energy is slowly decaying.

\begin{figure}
\includegraphics{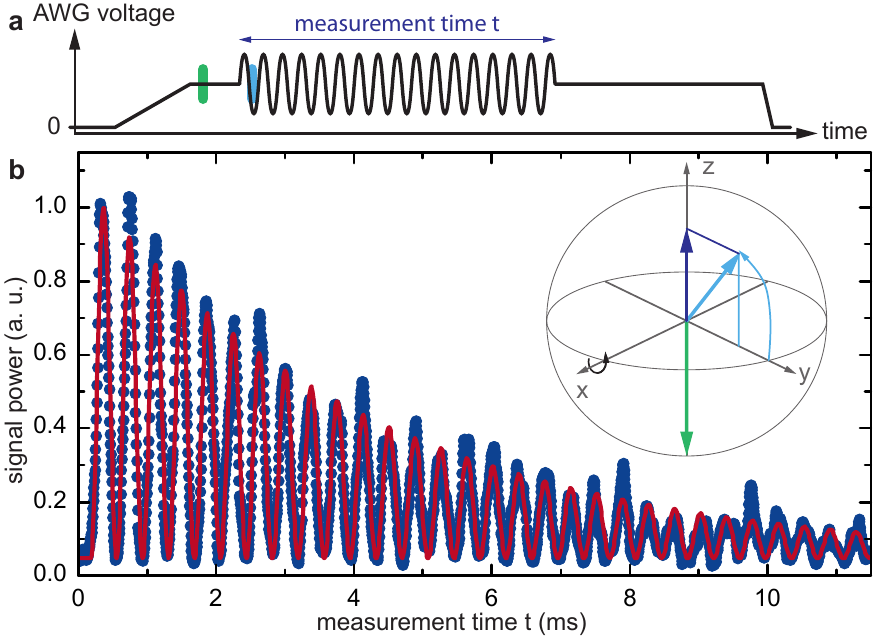}
\caption{\label{rabi}\bf Rabi oscillations. a\sf , Pulse scheme: the system is adiabatically tuned from the initialisation to the lower state, then a constant drive with frequency $\Omega$ is turned on. \bf b\sf , The z projections of the decaying Rabi oscillations (data: dark blue; fit: red) can be directly measured with a spectrum analyser. The Bloch sphere in the inset shows the state of the Bloch vector at selected times, which are marked in the same colour in \bf a\sf.
}
\end{figure}

Now, the application of a continuous pump tone with frequency $\Omega$ will start Rabi oscillations\,\cite{bib:Vandersypen2005,bib:Rabi1937} between the lower and upper state, as shown in \fig{rabi}.
They can be measured directly by monitoring the time evolution of the output power spectrum at the frequency of one of the hybrid modes, here shown for the upper state at 7.6028\,MHz, and measured with a bandwidth of 10\,kHz.
All time-resolved measurements are averaged over 20 (Rabi oscillations and T1 measurement) or 10 pulse sequences (Ramsey fringes and Hahn echo).
For a drive amplitude of 100\,mV (half peak-to-peak) we find a Rabi frequency of 8.3\,kHz (see section II.A of the Supplementary Information for the frequency dependence of the Rabi oscillations).
In principle, the decay of these oscillations is governed by both energy relaxation, characterised by a rate $1/T_1$, and phase decoherence, characterised by $1/T_2$ or $1/T_2^\ast$, where $T_2^\ast\leq T_2$ includes reversible processes caused by slow fluctuations or spatial inhomogeneity of the coupling.
For clarity, we use these well-known phenomenological constants in the same way as, e. g., in spin systems\,\cite{bib:Vandersypen2005}, as discussed in more detail in the Supplementary Information section I.

\begin{figure}
\includegraphics{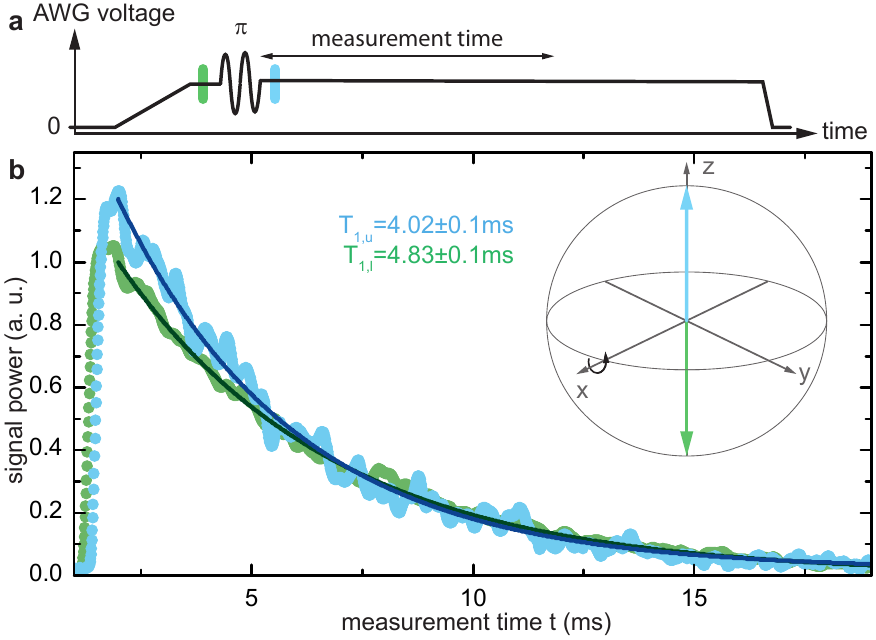}
\caption{\label{t1}\bf Energy relaxation. a\sf , Pulse scheme: the system is adiabatically tuned from the initialisation to the lower state. An additional $\pi$-pulse is used to rotate it to the upper state. \bf b\sf , Measured exponential decay of the lower (data: green; fit: dark green) and upper (data: blue; fit: dark blue) state. The Bloch sphere in the inset shows the state of the Bloch vector at selected times, which are marked in the same colour in \bf a\sf.
}
\end{figure}

The exponential decay of a state's energy defines $T_1$.
The corresponding measurement is shown in \fig{t1} for both lower and upper state: The system is once again prepared in the lower hybrid state.
To reach the upper state, a subsequent $\pi$-pulse is applied, thus performing one half of a Rabi cycle which transfers the system to the upper state (see Supplementary Information section II for details on the frequency and amplitude calibration of the applied pulses).
The exponential decay is then measured directly with a spectrum analyser at a bandwidth of 3\,kHz, exhibiting different relaxation times $T_{1,l}=4.83\pm 0.1\,$ms and $T_{1,u}=4.02\pm 0.1\,$ms for the lower and upper mode, respectively.
They correspond to the spectrally measured quality factors.
Previously, it has been shown that, at maximum coupling, the two hybrid modes should have the same quality factor and thus $T_1$ time\,\cite{Faust2012a}.
However, both modes are affected by dielectric damping\,\cite{rieger:103110}, leading to the observed difference.

\begin{figure}
\includegraphics{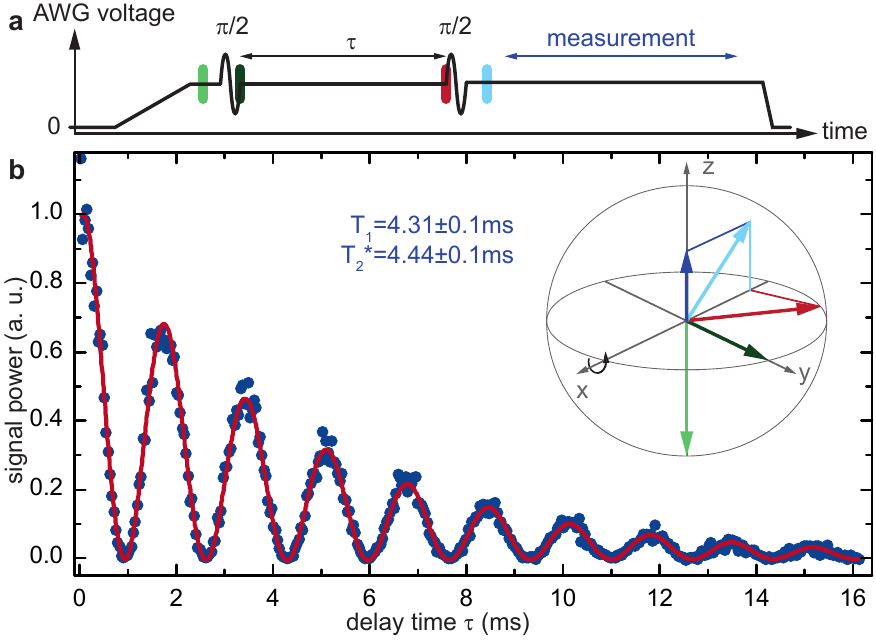}
\caption{\label{ramsey}\bf Ramsey  fringes. a\sf , Pulse scheme: the system is adiabatically tuned from the initialisation to the lower state. A $\pi/2$-pulse creates a superposition state, and after a delay $\tau$ a second $\pi/2$-pulse is applied. \bf b\sf , A 500\,Hz detuning between drive and precession frequency leads to a slow rotation of the superpostion state in the equator plane of the Bloch sphere, giving rise to a beating pattern in the measured z component after the second pulse (data: dark blue; fit: red). The Bloch sphere in the inset shows the state of the Bloch vector at selected times, which are marked in the same colour in \bf a\sf.
}
\end{figure}

To measure the $T_2^\ast$ time, a $\pi/2$-pulse is used after the preparation in the lower state to bring the system into a superposition state between lower and upper hybrid mode.
The frequency of the pulse is detuned to $\Omega+500\,$Hz, leading to a slow precession of the state vector around the z-axis of the Bloch sphere\,\cite{bib:Vandersypen2005,bib:Ramsey1950}.
As a result, a second $\pi/2$-pulse after time $\tau$ does not always bring the system into the upper state, but a slow oscillation, the so-called Ramsey fringes, is observed when the delay $\tau$ between the two pulses is varied and the z-projection of the state vector is measured after the second pulse, as shown in \fig{ramsey}.
The decay constant of this oscillation is $T_2^\ast$, while the decay of the mean value corresponds to an effective $T_1$ of both modes.
The fit in \fig[b]{ramsey} results in $T_2^\ast=4.44\pm0.1$\,ms and $T_1=4.31\pm0.1$\,ms.
The energy relaxation time of the superposition state $T_1$ is identical to the reciprocal rate average of the two hybrid modes 
\begin{equation*}
\overline{T_{1}}=2\left(\frac{1}{T_{1,l}}+\frac{1}{T_{1,u}}\right)^{-1}=4.39\,{\rm ms},
\end{equation*}
as the mechanical energy oscillates between the two modes with frequency $\Omega$ (see Supplemental Video).

\begin{figure}
\includegraphics{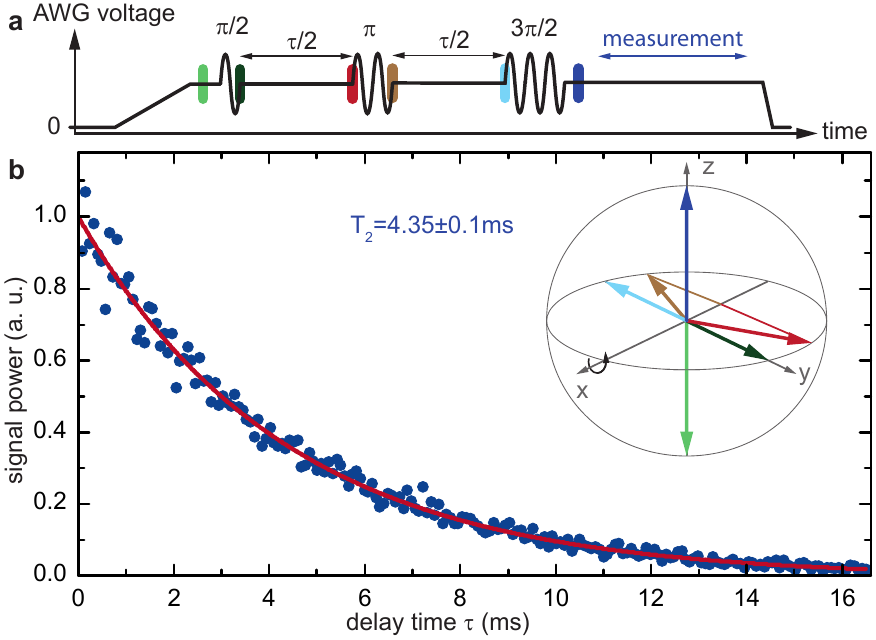}
\caption{\label{hahn}\bf Hahn echo. a\sf , Pulse scheme: the system is adiabatically tuned from the initialisation to the lower state. A $\pi/2$-pulse creates a superposition state, and after a delay $\tau/2$ a $\pi$-pulse mirrors the state vector to the other half of the Bloch sphere. After another delay of  $\tau/2$ a $3\pi/2$-pulse is used to rotate to the upper state. \bf b\sf , The inverse evolution of the system during the two delay times cancels out any broadening or slow precession effects, thus the system always ends up along the z-axis and no oscillation is observed (data: dark blue; fit: red). The Bloch sphere in the inset shows the state of the Bloch vector at selected times, which are marked in the same colour in \bf a\sf.
}
\end{figure}

By including an additional $\pi$-pulse at $\tau/2$ into the Ramsey pulse scheme and replacing the final $\pi/2$-pulse by an $3\pi/2$-pulse to once again rotate to the upper state (see \fig{hahn}), the $T_2$ time can be measured in a Hahn echo experiment\,\cite{bib:Vandersypen2005,bib:Hahn1950}.
The $180^\circ$ rotation flips the state vector in the xy-plane of the Bloch sphere, thus reversing the effects of a fluctuating or inhomogeneous coupling strength $\Omega$ in the second delay interval of $\tau/2$ and thereby canceling their contribution.
The frequency of the pulses is once again exactly $\Omega$, as all three pulses need to be applied exactly around the same axis.
The resulting decay curve represents $T_2$, for which a value of $T_2=4.35\pm0.1$\,ms can be extracted from the fit in \fig[b]{hahn}.

The good agreement between $T_2$ and $T_2^\ast$ clearly shows that reversible elastic dephasing, e. g. caused by temporal and spatial enviromental fluctuations or spatial inhomogeneities, does not noticeably increase decoherence.
Although the experiment is performed with billions of phonons, they all reside in the same collective mechanical mode and thus all experience an identical environment.
This strongly constrasts the behaviour found e. g. in spin qubits, where the hyperfine interaction with energetically degenerate nuclear spins causes $T_2^\ast\ll T_2$ (see Supplemental Information section III).

It is more surprising that the phase coherence time $T_2$ is equal to the average energy relaxation time $T_1$.
This indicates the absence of measurable elastic phase relaxation processes in the nanomechanical system, such that the observed loss of coherence is essentially caused by the energy decay of the mechanical oscillation\,(see also Supplemental Information section III).
Earlier research\,\cite{PhysRevLett.105.027205} suggests that the dominant relaxation mechanism in silicon nitride strings is mediated by localised defect states of the amorphous resonator material, described as two-level systems at low temperature.
They facilitate energy relaxation by providing the momentum required to transform a resonator phonon into a bulk phonon.
For this process to lead to elastic phase relaxation, an excited defect state would have to re-emit the phonon back into the resonator mode, which is extremely unlikely due to the weak coupling between the two.
In conclusion, we demonstrate coherent electrical control of a strongly coupled ($\Omega \gg \frac{f}{Q}$) nanomechanical two-level system, employing the pulse techniques well-known from coherent spin dynamics in the field of nanomechanics.
Each superposition state of the two hybrid modes on the Bloch sphere can be addressed by a sequence of the described pulses.
The presented system stands out by the finding that the elastic phase relaxation rate $\Gamma_\varphi$ is negligible compared to the energy decay rate $\frac{2\pi f}{Q}$, leaving room for improvement of the coherence via increased quality factors.

In light of the recent breakthrough in ground-state cooling of nanomechanical resonators\,\cite{OConnell2010,Teufel2011,Chan2011,PhysRevLett.108.033602}, the coherent manipulation schemes presented here open new applications for nanomechanical systems in quantum information.
Not only can they be used as efficient interfaces for quantum state transfers in hybrid quantum systems\,\cite{bib:Stannigel2010,bib:Meystre2012}, but by creating coupled, quantised resonators\,\cite{Brown2011} quantum computations can be carried out directly using nanoelectromechanical two-level systems\,\cite{2012arXiv1211.4456R}.

\section*{Acknowledgements}

Financial support by the Deutsche Forschungsgemeinschaft via Project No. Ko 416/18, the German Excellence Initiative via the Nanosystems Initiative Munich (NIM) and LMUexcellent, as well as the European Commission under the FET-Open project QNEMS (233992) is gratefully acknowledged.
We thank G.~Burkard for his comments on decoherence in a three-level system and H.~Okamoto, I.~Mahboob and H.~Yamaguchi for critically reading the manuscript.

\section*{Competing Interests}

The authors declare that they have no competing financial interests.\newline
\\
\\

\section*{Author Contributions}

J.R. and M.J.S. designed and fabricated the sample, T.F. conducted the measurements and analysed the data. T.F., J.P.K. and E.M.W. wrote the paper with input from the other authors, the results were discussed by all authors.

\section*{Correspondence}

Correspondence and requests for materials should be addressed to E.M.W.~(email: weig@lmu.de).

\bibliography{../coherence}

\renewcommand{\thefigure}{S\arabic{figure}}
 \renewcommand{\theequation}{S\arabic{equation}}
  \setcounter{figure}{0}
\setcounter{equation}{0}
\renewcommand*{\theHfigure}{\thepart.\thefigure}
\renewcommand*{\theHequation}{\thepart.\theequation}
\newcommand{\eq}[1]{equation~(\ref{#1})}
\clearpage

\section*{Supplemental Material to "Coherent control of a nanomechanical two-level system"}

\setcounter{section}{0}
\section{Relaxation times}

\begin{figure}[htb]
\includegraphics{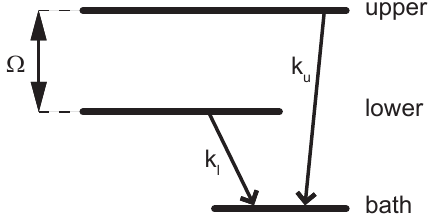}
\caption{\label{lscheme}\textbf{Levels of the mechanical system:} Schematic representation of the phonon bath and the upper and lower hybrid state of the mechanical system, separated by the frequency splitting $\Omega$, with respective decay constants $k_u$ and $k_l$.}
\end{figure}

In the main text, the relaxation constants $T_1$ and $T_2$ are introduced phenomenologically to define the exponential decay times extracted from the energy relaxation and Hahn echo experiments.
But different from the two-state spin systems associated with experiments on the Bloch sphere, the mechanical system investigated here actually has three states: the two coupled hybrid modes, i. e. the lower and upper state as well as the ground state (the thermally occupied phonon bath) into which a phonon can relax from either mode, see \fig{lscheme}.

For one, this makes it necessary to prepare the lower state prior to any measurement, as none of the states of the nanomechanical two-level system is automatically populated (except for the comparatively weak thermal excitation).
Furthermore, it introduces additional terms to the Bloch equations:
Assuming two independent decay rates $k_u$ and $k_l$ for the upper and lower state and following reference\,$^{31}$, the rotating-frame Bloch equations can be written as
\begin{align}
\label{Mx}\dot{M_x}(t)&=-\underbrace{\left(\frac{1}{T_\bot}+\overline{k}\right)}_{1/T_2}M_x(t)+\Delta M_y(t) \\
\label{My}\dot{M_y}(t)&=-\underbrace{\left(\frac{1}{T_\bot}+\overline{k}\right)}_{1/T_2}M_y(t)-\Delta M_x(t)-\omega_R M_z(t) \\
\label{Mz}\dot{M_z}(t)&=-\underbrace{\left(\frac{1}{T_\|}+\overline{k}\right)}_{1/T_1}M_z(t)+\omega_R M_y(t)
\end{align}
Here, $\Delta$ is the detuning between drive frequency and coupling strength $\Omega$ and $\omega_R$ reflects the drive strength and corresponds to the frequency of Rabi oscillations. $\overline{k}=\frac{k_u+k_l}{2}$ is the average decay rate, $T_\bot$ the relaxation time in the equator plane of the Bloch sphere, $T_\|$ the relaxation time along the z direction and $M_x$, $M_y$ and $M_z$ denote the respective components of the state vector.
The phenomenological parameters $1/T_1$ and $1/T_2$ can be identified as the sum of the respective rates.

The measured values of $T_1$ are consistent with the quality factors of the corresponding modes, which are only limited by $k_u$ and $k_l$.
This implies a negligibly small $1/T_\|$.
As the measurements show that $T_1$ is equal to $T_2$, $1/T_\bot$ must also be negligible.
Thus, the two coherence times $T_1$ and $T_2$ are solely limited by the average mechanical damping of the two resonator modes.

\begin{figure}
\includegraphics{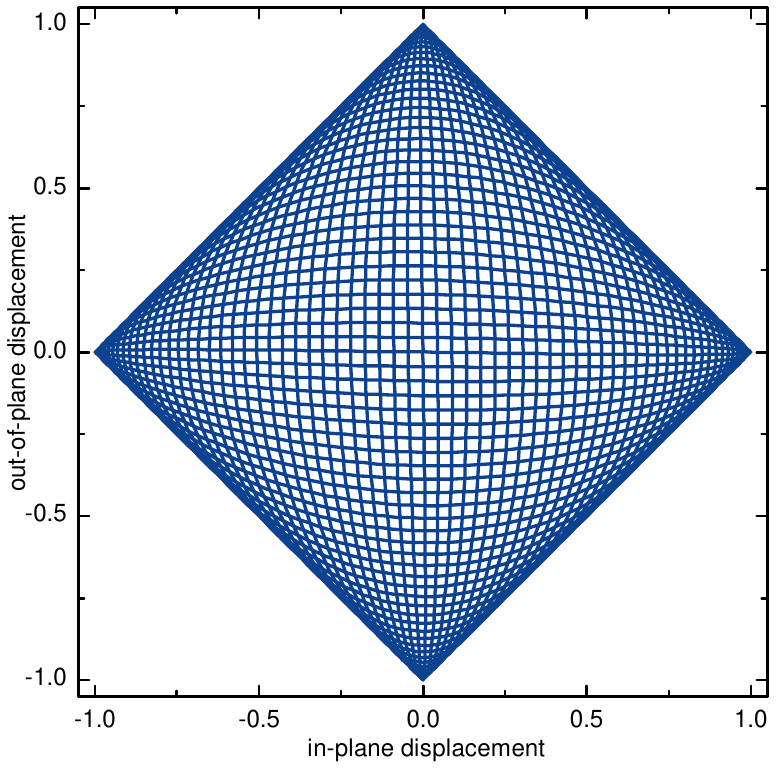}
\caption{\label{pspace}\textbf{Oscillation in the superposition state:} The normalized displacement of the resonator is calculated for one full precession period in the superposition state, i. e. for equal amplitudes of the lower and upper state.
This phase trajectory is plotted in the basis of the in-plane and out-of-plane mode (horizontal and vertical axis of the plot).}
\end{figure}

Reference 31 also introduces an additional term in \eq{Mz} taking into account the difference between the two decay rates $k_u$ and $k_l$.
It leads to a slow tilt of the state vector of a superposition state away from the equator plane towards the state with the smaller decay rate.
However, this effect plays no role in the classical system presented here: In a classical superposition state, the energy is distributed between both modes.
The system performs an oscillation between its two fundamental modes, changing the direction of rotation with the precession frequency.
Thus, the time spent in each mode is equal, and the state just experiences the average decay constant.
A plot of the phase space trajectory for one precession cycle is shown in \fig{pspace}, using a 10 times exaggerated coupling strength.
The horizontal and vertical axes correspond to an in-plane and out-of-plane oscillation, while a diagonal motion is associated with the lower and upper hybrid mode.
An animated version of this plot is available as an ancillary file on arXiv.org.

\section{Pulse calibration}

The first step in characterising the system is to measure an avoided crossing as shown in Figure~1 of the main text.
From its fit, the approximate frequency splitting $\Omega$ and the AWG voltage required to adiabatically tune to the lower state is extracted\,$^{9}$.

\subsection{Pulse frequency}

\begin{figure}[b]
\includegraphics{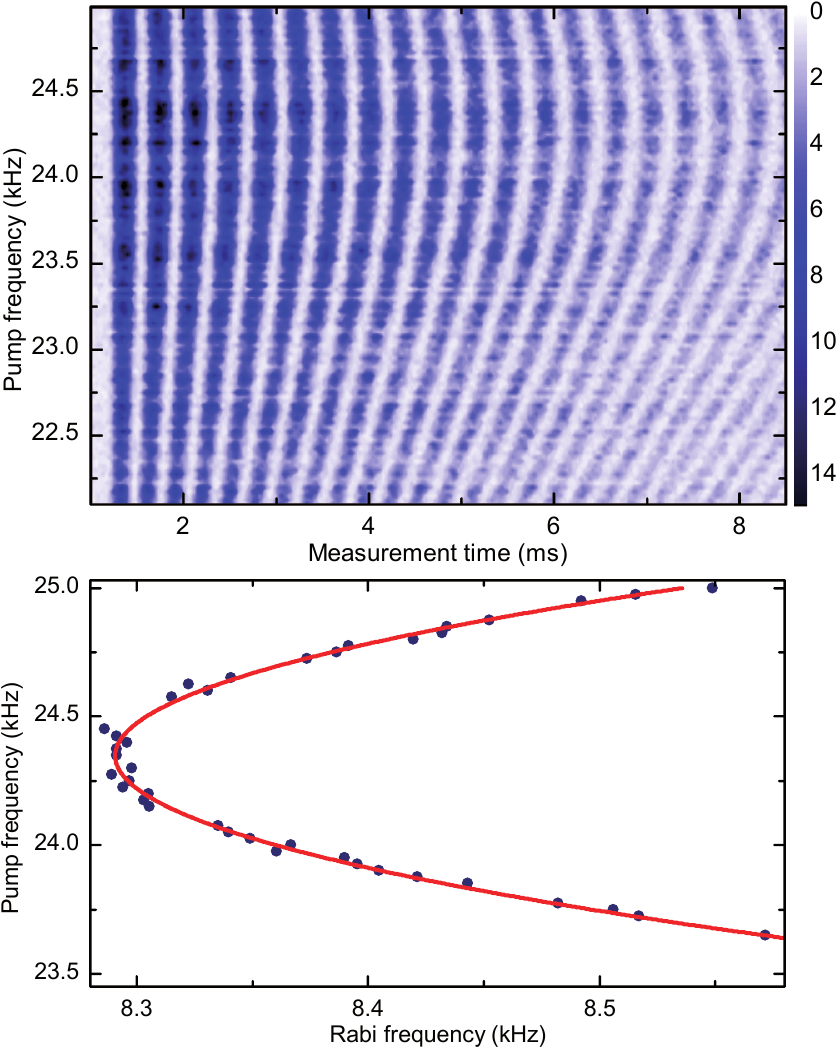}
\caption{\label{rsweep}\textbf{Pump frequency tuning:} The frequency of Rabi oscillations depends quadratically on the pump frequency detuning. The upper plot displays the measured signal power colour-coded versus pump frequency and measurement time. The extracted Rabi frequencies (blue points: data; red line: parabolic fit) are shown in the lower plot. The minimum of the parabola corresponds to the zero-detuning pump frequency.}
\end{figure}

To precisely determine the correct pump tone, the fitted frequency splitting is not accurate enough. Instead, the frequency of Rabi oscillations is monitored while sweeping the pump frequency.
The quadratic dependence for small detunings\,$^{2}$, as shown in \fig{rsweep}, allows to fit the measured points and extract the lowest Rabi frequency and thus the exact pump frequency corresponding to zero detuning.

\begin{figure}[b]
\includegraphics{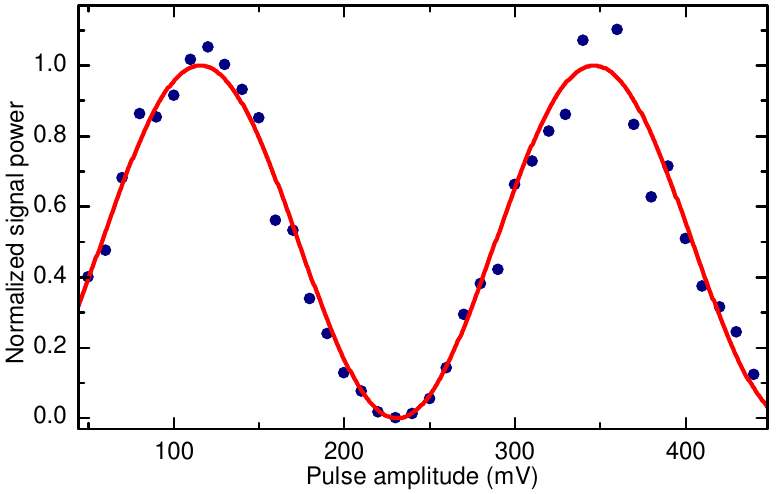}
\caption{\label{asweep}\textbf{Pulse amplitude sweep:} The population of the upper state after quadruple-sine-wave pulses of different amplitudes (blue points: data; red line: fit) demonstrates the correct behaviour for an amplitude of 231\,mV.}
\end{figure}

\subsection{Pulse length and spacing}

The pulses applied to the system should be as short as possible to allow fast control sequences.
The lower limit of the pulse length is one period of the pump signal, as abrupt voltage jumps of a chopped up sine wave will disturb the system.
The shortest applied pulse (a $\pi/2$ pulse) is thus set to a duration of 1/(pump frequency) so that it consists of exactly one sine wave. The other pulses are correspondingly longer, as shown in the pulse schemes in Figures~3-5 of the main text.

The delay time between two pulses in Figure~4 and 5 also needs to be a multiple of 1/(pump frequency), as otherwise the second rotation will not be carried out along the same axis.
This is apparent from the Ramsey fringe experiment, where the observed beating pattern, and thus the effective rotation angle, is caused by an intentional detuning of 500\,Hz.

\subsection{Pulse amplitude}

After both length and frequency of a pulse are fixed as described above, its amplitude has to be adjusted to define the rotation angle achieved with each pulse.
To this end, four sine periods (i. e. a $4\cdot\pi/2=2\pi$ pulse at the desired amplitude) are applied to the system, varying the amplitude.
The population of the upper state is measured.
At zero amplitude, the pulse has no effect and all energy remains in the lower state.
With increasing amplitude, the achieved rotation angle increases, and the first minimum corresponds to the desired $2\pi$ character of the pulse.
The population $P_u$ of the upper state can be described as
\begin{align}
P_u=\sin\left(\frac{\pi A}{A_0}\right)^2,
\end{align}
where A is the pulse amplitude and $A_0$ the amplitude corresponding to a rotation of $2\pi$.
A fit to the measured data, as shown in \fig{asweep}, can be used to extract the correct pulse amplitude.
This measurement has to be repeated with the detuned pump frequency used for the Ramsey fringes experiment, as the slightly different frequency also leads to a small shift in the amplitude of the pulse.

\section{Phase relaxation}

To define the elastic phase relaxation rate $\Gamma_\varphi$ two situations have to be distinguished: 
In systems where energy relaxation occurs only from the upper to the lower level (e.g. references\,$^{8, 32}$) the phase relaxation rate is defined as$^{31,33,34}$
\begin{equation}
\label{gp1}
\Gamma_\varphi=\frac{1}{T_2}-\frac{1}{2T_1}.
\end{equation}
In contrast to that, in a system dominated by spontaneous decay to a third state and no relaxation between the upper and lower level\,$^{33,34}$ (i. e. $1/T_\|=0$ and finite $\overline{k}$, see \eq{Mz}), the phase relaxation rate $\Gamma_\varphi$, in this case equal to $1/T_\bot$\,(defined in \eq{Mx} and \eq{My}), is given as
\begin{equation}
\label{gp2}
\Gamma_\varphi=\frac{1}{T_2}-\frac{1}{T_1}.
\end{equation}
This is the case for the system presented here.
As $T_1$ and $T_2$ are equal within the measurement acurracy, $\Gamma_\varphi$ can not be determined from the experiment.
This shows that the measured phase decoherence is solely caused by energy relaxation.
Processes changing the phase but preserving the state's energy seem to play no role.

To compare this mechanical two-level system to other coherent systems, it helps to take a general look at the possible decay processes:
Inelastic processes in which energy is transfered to a thermal bath are irreversible.
They are directly represented in the $T_1$ time and also pose a limit to $T_2$ via the two above equations \ref{gp1} or \ref{gp2}.
Irreversible elastic interactions lead to a non-zero $\Gamma_\varphi$ and thus reduce $T_2$, whereas reversible phase decay processes can be measured and controlled e. g. by a Hahn echo experiment and only decrease $T_2^\ast$.

In most coherent nanoscale solid-state systems, the coupling to a fluctuating thermal bath of phonons, photons, two-level systems or (nuclear) spins leads to one or more of the above processes.
For example, superconducting qubits\,$^{35-37}$ suffer from flux, charge and photon noise.
In gate-defined spin qubits e. g. nuclear spin\,$^{38-40}$ and phonon\,$^{41}$ interactions limit the performance, whereas NV centres in diamond\,$^{42}$ couple to the surrounding nuclear spin bath.
In the amorphous dielectric system presented here, relevant loss mechanisms occur via defect states with a broad energy spectrum often associated with two-level systems\,$^{43}$, and the phononic environment, as long as no additional electronic noise is introduced via the measurement devices and tuning voltages.
As the mechanical modes under investigation are situated within the suspended beam, they effectively reside inside a phonon cavity and couple extremely weakly to the phonon bath of the bulk sample via the narrow clamping points\,$^{44}$.
The exchange of energy of the discrete long-wavelength resonator modes and the continous shorter-wavelength phonon spectrum of the beam is found to be mediated by the defects.
As only higher energy phonons with small wavelengths can effectively transmit energy through the clamps to the bath, scattering of thermally excited higher energy phonons at defect states in a three particle interaction is the most likely process. 
These inelastic processes destroy energy as well as phase and likely explain why we find $T_1=T_2$.

\section*{Supplementary References}

\newcounter{qcounter}
\begin{list}{\arabic{qcounter}.}{\usecounter{qcounter}}
\setcounter{qcounter}{1}
\item
\bibinfo{author}{Vandersypen, L. M.~K.} \& \bibinfo{author}{Chuang, I.~L.}
\newblock \bibinfo{title}{{NMR techniques for quantum control and
  computation}}.
\newblock \emph{\bibinfo{journal}{Rev. Mod. Phys.}}
  \textbf{\bibinfo{volume}{76}}, \bibinfo{pages}{1037--1069}
  (\bibinfo{year}{2005}).
  \setcounter{qcounter}{7}
\item \bibinfo{author}{Burkard, G.}, \bibinfo{author}{Koch, R.~H.} \&
  \bibinfo{author}{DiVincenzo, D.~P.}
\newblock \bibinfo{title}{Multilevel quantum description of decoherence in
  superconducting qubits}.
\newblock \emph{\bibinfo{journal}{Phys. Rev. B}} \textbf{\bibinfo{volume}{69}},
  \bibinfo{pages}{064503} (\bibinfo{year}{2004}).
\item \bibinfo{author}{Faust, T.} \emph{et~al.}
\newblock \bibinfo{title}{Nonadiabatic Dynamics of Two Strongly Coupled
  Nanomechanical Resonator Modes}.
\newblock \emph{\bibinfo{journal}{Phys. Rev. Lett.}}
  \textbf{\bibinfo{volume}{109}}, \bibinfo{pages}{037205}
  (\bibinfo{year}{2012}).
\setcounter{qcounter}{30}
\item \bibinfo{author}{P\"ottinger, J.} \& \bibinfo{author}{Lendi, K.}
\newblock \bibinfo{title}{Generalized Bloch equations for decaying systems}.
\newblock \emph{\bibinfo{journal}{Phys. Rev. A}} \textbf{\bibinfo{volume}{31}},
  \bibinfo{pages}{1299--1309} (\bibinfo{year}{1985}).
  \item
\bibinfo{author}{Hu, X.}, \bibinfo{author}{de~Sousa, R.} \&
  \bibinfo{author}{Sarma, S.~D.}
\newblock \bibinfo{title}{Decoherence and dephasing in spin-based solid state
  quantum computers}.
\newblock \emph{\bibinfo{journal}{Proceedings of the 7th International
  Symposium on Foundations of Quantum Mechanics in the Light of New Technology,
  eds.Yoshimasa A. Ono, K. Fujikawa und Kazuo Fujikawa, World Scientific (or:
  cond-mat/0108339)}} \bibinfo{pages}{3--11} (\bibinfo{year}{2002}).
\item
\bibinfo{editor}{Drake, G. W.~F.} (ed.) \emph{\bibinfo{title}{{Springer
  Handbook of Atomic, Molecular, and Optical Physics}}}, page 1004
  (\bibinfo{publisher}{Springer}, \bibinfo{year}{2006}).
  \item
\bibinfo{author}{Burkard, G.} \emph{\bibinfo{title}{{Private communication}}}
 (\bibinfo{year}{2012}).
  
\item
\bibinfo{author}{Houck, A.}, \bibinfo{author}{Koch, J.},
  \bibinfo{author}{Devoret, M.}, \bibinfo{author}{Girvin, S.} \&
  \bibinfo{author}{Schoelkopf, R.}
\newblock \bibinfo{title}{Life after charge noise: recent results with transmon
  qubits}.
\newblock \emph{\bibinfo{journal}{Quantum Information Processing}}
  \textbf{\bibinfo{volume}{8}}, \bibinfo{pages}{105--115}
  (\bibinfo{year}{2009}).

\item
\bibinfo{author}{McDermott, R.}
\newblock \bibinfo{title}{Materials origins of decoherence in superconducting
  qubits}.
\newblock \emph{\bibinfo{journal}{IEEE Transactions on Applied
  Superconductivity}} \textbf{\bibinfo{volume}{19}}, \bibinfo{pages}{2 -- 13}
  (\bibinfo{year}{2009}).

\item
\bibinfo{author}{Rigetti, C.} \emph{et~al.}
\newblock \bibinfo{title}{Superconducting qubit in a waveguide cavity with a
  coherence time approaching 0.1 ms}.
\newblock \emph{\bibinfo{journal}{Phys. Rev. B}} \textbf{\bibinfo{volume}{86}},
  \bibinfo{pages}{100506} (\bibinfo{year}{2012}).
  
\item
\bibinfo{author}{Petta, J.~R.} \emph{et~al.}
\newblock \bibinfo{title}{Coherent manipulation of coupled electron spins in
  semiconductor quantum dots}.
\newblock \emph{\bibinfo{journal}{Science}} \textbf{\bibinfo{volume}{309}},
  \bibinfo{pages}{2180--2184} (\bibinfo{year}{2005}).

\item
\bibinfo{author}{Reilly, D.~J.} \emph{et~al.}
\newblock \bibinfo{title}{Suppressing spin qubit dephasing by nuclear state
  preparation}.
\newblock \emph{\bibinfo{journal}{Science}} \textbf{\bibinfo{volume}{321}},
  \bibinfo{pages}{817--821} (\bibinfo{year}{2008}).

\item
\bibinfo{author}{Bluhm, H.} \emph{et~al.}
\newblock \bibinfo{title}{Dephasing time of GaAs electron-spin qubits coupled
  to a nuclear bath exceeding 200\,\textmu s}.
\newblock \emph{\bibinfo{journal}{Nat Phys}} \textbf{\bibinfo{volume}{7}},
  \bibinfo{pages}{109--113} (\bibinfo{year}{2011}).

\item
\bibinfo{author}{Hanson, R.}, \bibinfo{author}{Kouwenhoven, L.~P.},
  \bibinfo{author}{Petta, J.~R.}, \bibinfo{author}{Tarucha, S.} \&
  \bibinfo{author}{Vandersypen, L. M.~K.}
\newblock \bibinfo{title}{Spins in few-electron quantum dots}.
\newblock \emph{\bibinfo{journal}{Rev. Mod. Phys.}}
  \textbf{\bibinfo{volume}{79}}, \bibinfo{pages}{1217--1265}
  (\bibinfo{year}{2007}).
  
\item
\bibinfo{author}{Takahashi, S.}, \bibinfo{author}{Hanson, R.},
  \bibinfo{author}{van Tol, J.}, \bibinfo{author}{Sherwin, M.~S.} \&
  \bibinfo{author}{Awschalom, D.~D.}
\newblock \bibinfo{title}{Quenching spin decoherence in diamond through spin
  bath polarization}.
\newblock \emph{\bibinfo{journal}{Phys. Rev. Lett.}}
  \textbf{\bibinfo{volume}{101}}, \bibinfo{pages}{047601}
  (\bibinfo{year}{2008}).
  
\item
\bibinfo{author}{Pohl, R.~O.}, \bibinfo{author}{Liu, X.},
  \bibinfo{author}{Thompson, E.}
\newblock \bibinfo{title}{Low-temperature thermal conductivity and acoustic attenuation in amorphous solids}.
\newblock \emph{\bibinfo{journal}{Rev. Mod. Phys.}} \textbf{\bibinfo{volume}{74}},
  \bibinfo{pages}{991--1013} (\bibinfo{year}{2002}).  
  
\item
\bibinfo{author}{Cole, G.~D.}, \bibinfo{author}{Wilson-Rae, I.},
  \bibinfo{author}{Werbach, K.}, \bibinfo{author}{Vanner, M.~R.} \&
  \bibinfo{author}{Aspelmeyer, M.}
\newblock \bibinfo{title}{Phonon-tunnelling dissipation in mechanical
  resonators}.
\newblock \emph{\bibinfo{journal}{Nat Commun}} \textbf{\bibinfo{volume}{2}},
  \bibinfo{pages}{231} (\bibinfo{year}{2011}).
  \\
  \\

\end{list}  

\end{document}